# A Comparison of Coronal Mass Ejection Models with Observations for Two Large CMEs Detected During the Whole Heliosphere Interval

Chia-Hsien Lin [1, *] and James Chen [2]

[1] *Graduate Institute of Space Sciences, National Central University, Taoyuan, Taiwan, R.O.C.*
[2] *Naval Research Laboratory, Plasma Physics Division, Washington, DC, U.S.A.*

## ABSTRACT

Two major coronal mass ejections (CMEs) observed during the Whole Heliosphere Interval (WHI) are compared with the catastrophe (CA) and eruptive flux rope (EF) models. The objective is to test two distinct mechanisms for CMEs by modeling these well-observed CMEs and comparing predictions of the theories and observed data. The two CMEs selected for this study occurred on 25 March and 5 April 2008, respectively. For the 25 March event, an M 1.7 class flare, a filament eruption, and hard X-ray (HXR) and soft X-ray (SXR) emissions were observed during the CME onset. The observed CME kinematics and SXR light curve of this event are found to be more consistent with the EF model than with the CA model. For the 5 April event, the SXR light curve shows multiple enhancements, some of which temporally coincide with successive side loop brightening and multiple foot points at the source region after the eruption. The physical connection between the side-loop multiple brightenings and the eruption cannot be determined from the data. Both models produced observationally consistent kinematics profiles, and the EF model correctly predicted the first emission enhancement. Neither model includes multiple brightenings in the formulation.

## 1. INTRODUCTION

Coronal mass ejections (CMEs) are large-scale eruptive events from the Sun. They eject large amounts of mass and magnetic flux from the Sun into interplanetary space. These ejections are now recognized as the main solar source of major disturbances in space weather and can have severe impacts on Earth, such as damaging space instruments, disrupting communications and navigation systems, and causing outages of electric power grids. Developing accurate and timely forecasting methods for "geoeffective" (i.e., storm-causing) solar eruptions and their terrestrial consequences has been a great research interest. To date, the basic strategy to predict CMEs has been based on statistical studies of photospheric magnetic features in the source regions associated with eruptions. For instance, Canfield et al. (1999) examined active regions with different types of coronal loops, and concluded that active regions with "sigmoidal" coronal loops are more likely to erupt. Georgoulis (2008) investigated the magnetic complexity of active regions and found that CME speed is often correlated with the intensity of polarity inversion lines: faster CMEs are more likely to be launched from active regions with stronger and more concentrated bipolar magnetic flux. In addition to statistical studies based on observed properties, Kusano et al. (2012) used numerical simulation to investigate the dynamics of different magnetic structures and determined two magnetic configurations that are closely associated with solar eruptions. Despite these efforts, however, many studies have shown that current solar eruption forecasting methods are far from reliable, and significant improvement is needed (e.g., Petrie et al. 2011; Welsch et al. 2011).

The ability to forecast CMEs can be improved by a better understanding of the physics behind these eruptions. To study the possible mechanisms responsible for CME eruptions, numerous models have been developed over several decades (see Lin et al. 2003, for a review). These models can be divided into two categories: the ideal magnetohydrodynamics (MHD) type, in which no magnetic diffusivity or resistivity is included in the formulation and the resistive MHD type, in which non-ideal MHD processes, such as magnetic reconnection, play essential roles in the eruption process.

---

* Corresponding author
 E-mail: *chlin@jupiter.ss.ncu.edu.tw*

For this study, we have chosen two well-documented models, one from each type, and compare their theoretical predictions with specific CME events in a quantitative way. The objective is to examine whether the predictions of the two models are sufficiently distinct so that their level of agreement, or disagreement, with observations can be used to determine which model is the more probable mechanism for CMEs and whether the model assumptions are appropriate.

We followed the basic analysis method described in Lin et al. (2010) for a comparison between models and observations. Lin et al. (2010) selected three models for their study: Toroidal Instability model (TI) (Kliem and Török 2006), Catastrophe model (CA) (Priest and Forbes 1990; Forbes and Isenberg 1991; Forbes and Priest 1995; Lin and Forbes 2000; Priest and Forbes 2000), and Breakout model (BO) (Antiochos et al. 1999), among which TI is an ideal MHD model and CA and BO can both be categorized as the non-ideal MHD type. Lin et al. (2010) implemented an exponential function to represent the TI model. An empirical piece-wise polynomial function was used, that is, fitting different parts of the CME trajectory with different polynomials, to represent the BO model. However, the exponential function derived by Kliem and Török (2006) is only applicable during the early eruption stage and the piece-wise polynomial function does not contain the actual BO model physics. Such comparisons are not well constrained based on physics. [Also see Chen (2007) for some caveats concerning the TI model.]

In this study, we conducted a better constrained and more quantitative analysis by focusing on models in which theoretical solutions can be obtained for specific events. For this reason we replaced the TI model with the eruptive flux rope (EF) model (Chen 1989), which is an ideal MHD model. We can compute the entire theoretical eruption trajectory from initiation to the late propagation stage. We supplemented the CA model in Lin et al. (2010) with an estimation of the magnetic reconnection rate. The BO model, which is computed using MHD simulations with no available theoretical kinematic profile, is not examined in this study.

Our main strategy is to compare the theoretical CME trajectories of the two different models for the observed events with the observationally determined profiles. Since these theoretical trajectories are the mathematical solutions to a set of equations of motion relating physical quantities, we must identify observables that are directly comparable to the calculated quantities, which serve as the physical constraint on the solutions. Furthermore, once a solution most consistent with the observed quantities is obtained, additional theoretical predictions of the solution for other quantities must be derived and compared with observed values to validate consistency. For this study, we will calculate theoretical solutions to fit observed CME trajectories (height-time data). The best solutions will be used to generate the temporal profile of observed X-ray data. The two models represent two distinct processes relating the CME acceleration to the X-ray emission profiles. In addition to the quantitative comparison, various observed features and events associated with eruptions will be qualitatively compared with the predicted scenarios from different models to further test consistency. Note that by "most consistent" with the observed data, we refer to comparison with the height-time data throughout the available field of view.

The CMEs selected for comparison are the two most prominent CMEs observed during the Whole Heliosphere Interval (WHI). This was an internationally coordinated effort to study the 3D interconnected solar-heliospheric-planetary system. The two CMEs occurred on 25 March and 5 April 2008, respectively. While the 25 March event has been studied by several authors (e.g., Chen and Kunkel 2010; Temmer et al. 2010), the trajectory of the CME in these studies was derived from observation from a single spacecraft, either STEREO/SECCHI-A or -B (Solar TErrestrial RElations Observatory/Sun Earth Connection Coronal and Heliospheric Investigation) satellite. The direction of CME propagation was assumed on the basis of the source region location, and the projection effects were not specifically addressed. We reconstructed the 3D trajectory of the event using images from twin instruments A and B of the STEREO/SECCHI satellites (Howard et al. 2008), which helps reduce the uncertainties about the 3D geometry of these structures and the propagation direction.

The theoretical basis for the two models and their similarities and differences are discussed in section 2. We do not purport to provide a comprehensive study of every aspect of these models, but, instead, will focus on a number of specific, well-observed properties, primarily the CME height-time data and Geostationary Operational Environmental Satellite (GOES) SXR data as well as the candidate source region configurations. The data and analysis procedures are described in section 3. Our analysis results are presented and discussed in section 4. Major conclusions of this work are summarized in section 5.

## 2. CME MODELS
### 2.1 CA

The CA model configuration consists of a two dimensional magnetic flux rope initially in equilibrium. The flux rope is generated by placing two linear sources of magnetic field with opposite polarities on the surface of the Sun. An illustration of the model is shown in Fig. 1. The model proposes that the flux rope equilibrium can be destroyed by quasi-statically moving the two line sources toward each other and that when the distance between the line sources become shorter than a critical distance, the upward magnetic pressure force would overtake the downward tension force, dynamically propelling the flux rope upward. This eruptive process is termed a catastrophic loss of equilibrium. The initial eruptive process is in the realm of ideal MHD. After the ideal MHD evolution, the flux rope dynamics are critically determined by the formation of current sheets (CSs) and magnetic reconnection. The model shows that in order for the erupting flux rope to escape the solar surface, magnetic reconnection rate must be sufficiently high; otherwise, the flux rope would stop rising once it reaches a new equilibrium and its kinetic energy is dissipated. To allow magnetic reconnection, a CS, or X-line, must form below the erupting flux rope. The field lines inside the CS are no longer frozen to the plasma but can move through each other.

The computation of the full model is divided into multiple stages with the initial ideal MHD stage followed by a non-ideal phase during which the topology of the arcade changes via reconnection and a CS is formed (Lin and Forbes 2000). For the ideal MHD initial eruption process, Priest and Forbes (2000) derived an analytical expression for a thin flux-rope, where the flux-rope radius can be considered to be much smaller than a characteristic height:

$$\dot{h} \approx \sqrt{\frac{8}{\pi}\left[\ln\left(\frac{h}{\lambda_0}\right) + \frac{\pi}{2} - 2\tan^{-1}\left(\frac{h}{\lambda_0}\right)\right]^{1/2}} + \dot{h}_0 \quad (1)$$

where $h$ is the height of the flux rope, $h_0$ is the initial height, and $\lambda_0$, called critical height, is the height of the flux rope when the loss of balance occurs.

After the ideal MHD initial eruption, a CS is formed and magnetic reconnection is prescribed by the model to occur at the mid-point of the CS with a uniform reconnection rate along the sheet. The reconnection rate is specified as the rate of magnetic flux crossing the reconnection point related to the electric field in the CS according to Faraday's law: (e.g., Lin and Forbes 2000):

$$\begin{aligned}\text{Reconnection Rate} &= -\frac{1}{c}\frac{\partial A(0, y_0)}{\partial t} \\ &= E_z = M_A V_A(0, y_0) B_y(0, y_0)/c\end{aligned} \quad (2)$$

where $y_0$ is the reconnection point, $A(0, y_0)$ is the magnetic flux passing through the CS, $E_z$ is the electric field in the CS, $V_A$ is the Alfvén speed, and $B_y$ is the magnetic field along the CS. Here, $M_A \equiv |V_R/V_A|$ is the Alfvén Mach number, with $V_R$ the velocity of the plasma flowing into the CS. $V_A$ and $B_y$ at the reconnection point are prescribed for each configuration under consideration, so that the reconnection rate is essentially parameterized by $M_A$. Different flux-rope motion can be produced by tuning $M_A$. Lin and Forbes (2000) varied $M_A$ to examine the effect of the reconnection rate on the CME motion. They found that if $M_A$ is not greater than 0.001, the flux rope would oscillate around an equilibrium height until the downward and upward forces balance each other again. For an intermediate reconnection rate, $0.001 < M_A < 0.041$, the CME would undergo a brief deceleration phase after the initial loss of balance, and then accelerate again to reach peak acceleration. If $M_A$ is greater than 0.041, the deceleration phase does not occur. It is noted that at this time, the initial ideal MHD evolution and the subsequent magnetic reconnection are treated separately in that the transition is not self-consistently determined within the model.

## 2.2 EF Model

The EF model configuration is a 3D flux rope embedded in an ambient coronal magnetic field ($B_c$) which, after reaching the peak value just above the initial position of the flux rope apex, decreases exponentially with height. A flux rope is defined as a system of electric current and its magnetic field, both of which have toroidal and poloidal components. An illustration of the model is shown in Fig. 2. The entire flux rope dynamics are treated within the ideal MHD framework. A detailed description of the model and derivation of the equations can be found in previously published papers (e.g., Chen 1989, 1996; Chen and Kunkel 2010). Here, we briefly review the basic principles and equations relevant to our analysis.

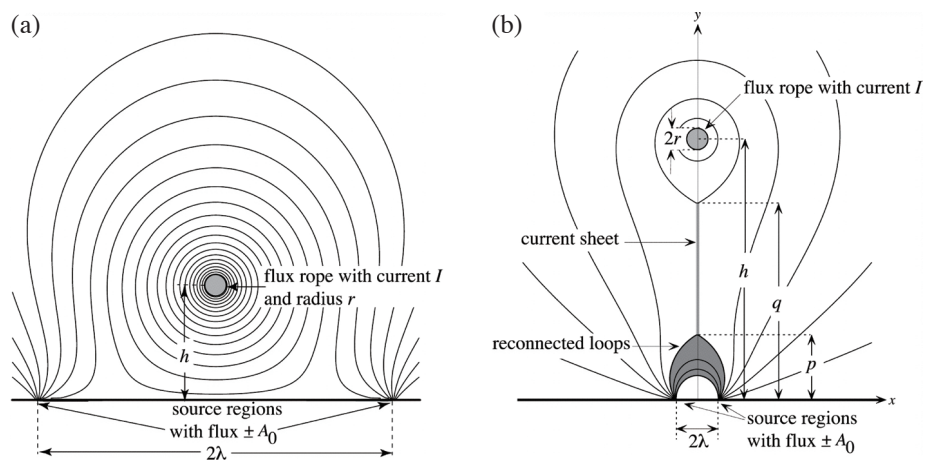

Fig. 1. The 2D illustration of the catastrophe model (CA) (Reeves 2006). (a) Magnetic configuration before the initial eruption. (b) Magnetic configuration after the initial eruption and current sheet (CS) formation.

The flux rope motion is governed mainly by the Lorentz force, which is composed of an outward $J_t B_p$ force and a downward $J_p B_t$ force, where $J_t$ and $J_p$ are the toroidal and poloidal components of the currents and $B_t$ and $B_p$ the toroidal and poloidal components of the magnetic field of the flux rope. The basic idea of the model is that because of the flux rope curvature, the Lorentz force per unit length has a net force in the major radial direction. The equation governing the motion of the centroid of the flux rope apex can be written as (Chen and Kunkel 2010):

$$M\frac{d^2 Z_{ce}}{dt^2} = \frac{\Phi_p^2}{c^4 L^2 R}\left[\frac{\ln\left(\frac{8R}{a}\right) - 1 + \frac{\xi_i}{2} - \frac{1}{2}\frac{\overline{B}_t^2}{B_{pa}^2} +}{\frac{1}{2}\beta_p + 2\left(\frac{R}{a}\right)\frac{B_c(Z_{ce})}{B_{pa}}}\right] + F_g + F_d \quad (3)$$

Where the symbols are defined as follows: $M$ is the mass per unit length; $Z_{ce}$ is the flux-rope centroid (ce) heigh from the photosphere; $\Phi_p$ is the poloidal magnetic flux; $L$ is the self-inductance; $R$ and $a$ are the major and minor radii; $\overline{B}_t$ is the toroidal magnetic field averaged over the minor radius; $B_{pa}$ is the poloidal field at $r = a$; $\beta_p$ is defined as $8\pi(\overline{p} - p_c)/B_{pa}^2$; where $\overline{p}$ is the average internal pressure and $p_c$ the ambient coronal pressure; $B_c$ is the ambient coronal field perpendicular to the flux rope; $\xi_i$ is the internal inductance; and $F_g$ and $F_d$ are the gravitational and drag forces. It is worth noting that the aforementioned outward and inward components of the Lorentz self-force correspond to $R^{-1}[\ln(8R/a) - 1 + \xi_i/2]$ and $-R^{-1}(\overline{B}_t^2/2B_{pa}^2)$ in Eq. (3), respectively. The trajectories of the leading edge (le) of the flux rope is denoted by $Z_{le}$, top of current channel (cr), and the prominence (pr) are derived by setting $Z_{le} = Z_{ce} + 2a$, $Z_{cr} = Z_{ce} + a$, and $Z_{pr} = Z_{ce} - a$, where $Z_{ce}$ is the height of the centroid of the apex. Since the minor radius $a$ also changes in time as the flux rope expands outward, the following equation is incorporated in the model calculation to account for the dynamics of the minor radius (Chen and Kunkel 2010):

$$M\frac{d^2 a}{dt^2} = \frac{I_t^2}{c^2 a}\left(\frac{\overline{B}_t^2}{B_{pa}^2} - 1 + \beta_p\right) \quad (4)$$

In the model, the flux rope is dynamically driven out of equilibrium by an increase in poloidal magnetic flux (i.e., $d\Phi_p/dt$). It should be pointed out that the physical source of $d\Phi_p/dt$ is not specified in the formulation: the source may be in the corona or below the photosphere. Krall et al. (2000) investigated different physical scenarios and the corresponding mathematical profiles for $d\Phi_p/dt$ - for example, a pressure impulse, gradual increase in the poloidal flux that may result from a slow *twisting* motion of magnetic footpoints, and increasing $\Phi_p(t)$ on the timescale of eruption - and found that the last scenario - increasing $\Phi_p(t)$ on the timescale of eruption - produced the best match to the characteristic acceleration and propagation profiles in the LASCO field of view. The temporal profiles of $d\Phi_p(t)/dt$ that produces the best match to the observed CME trajectories (height-time data) were later shown to be in agreement with the temporal profiles of soft X-ray (SXR) emissions from the flares associated with given CMEs (Chen and Kunkel 2010).

To explain CME-associated solar flares and SXR emissions, Chen and Kunkel (2010) proposed that the electromotive force (EMF), $\varepsilon(t) \equiv -(1/c)d\Phi_p(t)/dt$, which is necessarily generated by a non-zero $d\Phi_p(t)/dt$, would accelerate charged particles leading to the SXR emissions. They suggested that an observational indication of such mechanism would be a correlation between the temporal profile of SXR emission and $d\Phi_p(t)/dt$. As part of our theory validation study, we compare the SXR profile and $d\Phi_p(t)/dt$.

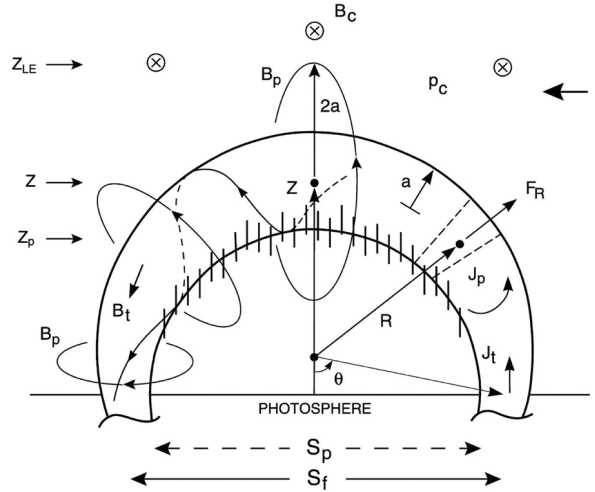

Fig. 2. Face-on representation of the eruptive flux rope (EF) model (Chen et al. 2006). The illustration shows a flux rope of major radius R and minor radius a embedded in an ambient corona with magnetic field $B_c$ and pressure $P_c$. The subscripts t and p refer to the toroidal and poloidal components of magnetic field B and current density J of the flux rope. Z is the apex height of the centroid of the flux rope, measured from the photosphere. The surface of the flux rope is located at 2a from the centroid. The CME leading edge is at $Z_{le} = Z + 2a$, the top of the current channel is at $Z_{cr} = Z + a$, and the prominence is located at $Z_p = Z - a$. The foot point separation of the flux rope is $S_f$ between the centers at the two ends. $S_p$ is the foot point separation of the prominence $S_p = S_f - 2a_f$, where $a_f$ is the minor radius at the foot point. $F_R$ is the net force in the radial direction.

## 2.3 Model Comparisons

In the CA model, the quantity $\partial A(0, y_0)/\partial t$ in Eq. (2) - the reconnection electric field - plays the equivalent role to $d\Phi_p/dt$ in the EF model. From Fig. 1b, we see that in order to satisfy magnetic flux conservation, an increase in the magnetic flux passing through the CS is equal to the increase of the magnetic flux surrounding the flux rope, which is the poloidal magnetic flux. Therefore, despite differences in the respective pictorial representations and the general belief that the two models are distinct, the mathematical representations have a number of specific commonalities, an important one being that the poloidal magnetic flux of the flux rope must increase at a sufficiently high rate in order for the flux rope to erupt and become a CME.

If a CME is driven by the CA mechanism, the initial ideal MHD expansion should start before the rise of SXR emission resulting from particle acceleration by the reconnection electric field and subsequent radiation, and the signatures of the magnetic reconnection should appear below the flux rope and after the eruption onset. For example, Priest and Forbes (2000) found in their simulation that the reconnection did not start until seven minutes after initiation. The theoretical kinematics derived from Eq. (1) should be consistent with the observationally determined kinematics during the initial eruption, and deviate from the observation after magnetic reconnection. When a deviation is observed, we utilize a computer code of the model (Lin and Forbes 2000, private communication) that estimates the reconnection rate as parameterized by $M_A$ for the model configuration. That is, any deviation from the ideal MHD trajectory is assumed to be due to reconnection.

To examine whether a CME can be driven by the EF mechanism, we compare the theoretical kinematics of the leading edge ($Z_{le}$) and the current channel top $Z_{cr}$ with the kinematics derived from the observed CME trajectory. The $d\Phi/dt$ profile predicted by the theoretical kinematics that is most consistent with the observation is subsequently compared with the observed SXR light curve.

## 3. DATA AND ANALYSIS

The two selected CMEs occurred on 25 March and 5 April 2008, respectively. Viewed from the Earth, the 25 March event was originated in AR10989 (S13E80) and the 5 April event in AR10987 (S8W115) (Nitta 2011). The CMEs are distinguished in the rest of this paper by the month in which they were observed, with "03" and "04" referring to March and April, respectively; that is, CME03 for the 25 March event and CME04 for the 5 April one. We utilized the images obtained from STEREO/SECCHI instruments (EUVI 171°A channel, COR1, and COR2) to derive the kinematics. We also incorporated the SXR data from GOES (Garcia 1994) 1 - 8°A channel, and hard X-ray (HXR) data from RHESSI [Reuven RAmaty High Energy Solar Spectroscopic Imager; Lin et al. (2002)] for the associated flare information.

For a qualitative comparison, we examined certain observable properties and effects related to the CMEs, such as the geometry and morphology of the source regions, SXR and HXR light curves during the process of eruption, and flares and/or filament eruptions associated with CMEs. These observed properties and sequence of effects were compared with the scenarios predicted by the two models to examine if the models are consistent with the observation. For the quantitative comparison, different theoretical CME trajectory profiles were matched to the observed CME trajectory. The solutions were then used to derive the "theoretical" velocity and acceleration, which were compared with the velocity and acceleration derived from the observed trajectory. The theoretical trajectory profiles implemented in the study are: solutions of Eq. (1) for the CA model, $Z_{ce} + 2a$ for the EF model LE (denoted EFle), and $Z_{ce} + a$ for leading edge of the current channel for the EF model (denoted by EFcr), where $Z_{ce}(t)$ and $a(t)$ are the centroid height from the photosphere and minor radius at the apex, respectively, which are determined by Eqs. (3) and (4).

## 4. RESULTS

### 4.1 25 March 2008 Event (CME03)

CME03 was observed by both SECCHI A and B instruments. We determined the trajectory of the CME front from the solar surface as follows: the CME front in each synchronized A- and B-image pair was traced simultaneously using the SolarSoft procedure scc_measure.pro. The coordinates of the tracked points were then used to calculate their true distances from the solar center. This procedure was repeated multiple times to cover the bright CME loop at every time step. The "observed" height of the CME LE was determined by taking the average of these measurements, and we used the measured thickness of the bright loop - the spread in the measured heights - as the observational uncertainty. This is typically ~1% of the measured height, we chose the larger of 1% of the measured height or the spread given by the multiple measurement. This is the method used to determine the position of any bright features. In essence, we use the centroid of a bright feature as its position, i.e., elongation, and its spread as the uncertain for the underlying magnetic structure. The velocity and acceleration were calculated using the time derivatives of the estimated height, and their respective errors were computed using the standard error propagation method.

CME03 was associated with an M 1.7 class flare as observed by GOES. The SXR emission of GOES 1 - 8°A channel showed a steep rise after 18:40 UT, and peaked at approximately 18:56 UT. An enhanced HXR emission in the RHESSI 18 - 30 keV energy band was detected from AR10989 during the eruption. According to the RHESSI flare list[1], the HXR emission started to rise at 18:44:24 UT and peaked at 18:51:58 UT. Temmer et al. (2010), however, determined from their analysis that the peak HXR emission was at 18:51:34 UT. The difference is 24 s, which can be considered as uncertainty in the timing of the HXR maximum. Figure 3 shows the initial eruption phase of this CME. The running-difference images of the EUVI-B and -A instruments are shown in the left-hand-side (LHS) and right-hand-side (RHS) columns respectively, and the observation times are as indicated above the corresponding panels. The red stars mark the CME LE. The EUVI images were enhanced using a wavelet image-processing technique (Stenborg et al. 2008). The bright feature at the CME core is an erupting prominence.

The observed profiles of the CME kinematics and SXR light curve are plotted in the LHS column of Fig. 4. The kinematics during the early stage of the CME can be viewed in more detail in the RHS column of Fig. 5. The plot (right, middle panel) shows that the speed of the first point (at ~18:45 UT) is greater than 300 km s$^{-1}$, indicating that the eruption began at an earlier time. By a linear extrapolation of the measured velocity points, we can estimate that the velocity rose from zero after ~18:40 UT. Temmer et al. (2010) first detected the flux rope at ~18:35 UT. Their derived kinematics indicate that the flux rope began to accelerate at ~18:40 UT (cf. Fig. 12 in their paper), which is approximately the same time as when the SXR emission began to increase, and is remarkably close to the value estimated by our simple extrapolation. This apparent agreement, however, should be taken only as showing consistency for this event as the simple extrapolation does not take into account the nontrivial acceleration and other dynamical properties that are generally present.

For a quantitative analysis, the theoretical solutions that best match the observed height-time data of a number of CME features have been obtained for the two models (CA for the catastrophe model; EFle and EFcr for the EF model). These are plotted in Fig. 5. The symbols are the observed data points and different continuous curves show the different solutions, as indicated in the plot. To allow detailed inspection of the complete kinematic profile and the dynamics during the early acceleration phase, the results for the entire data set are placed in the LHS column and those for the low corona part in the RHS column.

For the CA model, the best-match solution we found is reasonably consistent with the rise phase of the observed height-time and therefore acceleration-time profiles. The critical height $\lambda_0$ (i.e., the height of the flux rope when the balance is lost) and the Alfvén speed at this height inferred from this solution are $\lambda_0 \approx 131$ Mm and $V_A(\lambda_0) \approx 600$ km s$^{-1}$. These values correspond to a foot point separation of $\approx 262$ Mm, and an ambient magnetic field strength of $\approx 22$ G for a plasma density $\approx 10^{-4}$ kg km$^{-3}$.

The velocity of the CA solution, however, deviates from the data points after ~19:10 UT. Specifically, the slope of the velocity-time curve (Fig. 5, left, middle panel) begins to decrease significantly earlier than is indicated by the COR2 data after approximately 19:00 UT. Furthermore, the data show that the CME decelerated while the CA solution shows decreasing but continued acceleration throughout the COR2 field of view. Since the best-fit CA solution [Eq. (1)] was derived from two dimensional force-free ideal MHD equations to describe the ideal MHD part of the eruption process, one interpretation of this discrepancy within the framework of the CA model is that a magnetic reconnection process is necessary in order to continually accelerate the flux rope. Another marked discrepancy is that the solution shows continued acceleration throughout the field of view while the data show noticeable deceleration after approximately 19:30 UT. This means that the equation for the CA fit does not provide the deceleration mechanism implied by the data. Evidently, the force acting on the CME as described by simple Eq. (1) and that exhibited by the data (i.e., acceleration, or force per unit mass) are significantly different.

To obtain a qualitative estimate of the reconnection rate implied by the best-fit CA solution, we utilized a computation code developed by Lin and Forbes (2000) (private communication). The code was formulated to compute the kinematics and dynamics of the specific flux rope discussed in Lin and Forbes (2000). The plasma condition and the flux rope configuration are built-in to this computation code and non-adjustable. Therefore, the comparison between the computed profile and the observation is only qualitative. For our analysis, the source-region photospheric magnetic field strength in the code was set to 40 Gauss, which is approximately the average strong unsigned magnetic flux of AR10989 (Baldner et al. 2013, private communication). The Alfvén Mach number $M_A$ was then tuned until the computation produced a kinematic profile that qualitatively replicated the data. The best agreement was reached when $M_A \approx 0.06$. According to Lin and Forbes (2000), this value corresponds to a high reconnection rate.

In contrast to the deviation seen in the CA fit, the EF fit follows the observed data points from the beginning to the last point. According to the solution for the EFle dynamics, the initial height of the flux rope ($Z_0$) is approximately 90 Mm measured from the base of corona, and the footpoint separation for comparison, $S_f$, is $\approx 180$ Mm. The MDI magnetogram of AR10989 at the disk center (cf. left panel in Fig. 6) shows that the size of the region is approximately 200 arcsec ($\approx 150$ Mm). Since the flux-rope footpoints cannot be directly observed, we use this AR dimension as an observational proxy for the actual footpoint separation. The footpoint separation predicted from the EF-fit is approximately 20% greater than the estimate based on the observational proxy provided by the MDI data. This is consistent with the degree of uncertainty found in previous theory-data comparisons (Chen et al. 2006).

---

[1] http://hessi.ssl.berkeley.edu/hessidata/dbase/hessi_flare_list.txt

Comparison between EFle (blue dotted line) and EFcr (red dashed line) in the top right panel of Fig. 5 shows that EFcr matches the EUVI part of the data slightly better than EFle does, although overall the solution accurately captures the main acceleration phase and the subsequent deceleration. The slight difference noted here may be due a deviation of the observed CME minor cross-section from the idealized circular cross-section or that the LE identified in the EUVI 171°A images and in the COR1/COR2 white-light images may not correspond to the same part of the (unobservable) magnetic structure. The bottom right panel shows that the calculated acceleration profile of EFle (blue dotted line) peaks at ~18:50 UT, which coincides with the peak of the $dI_{SXR}/dt$ (solid line). The peak of the HXR emission (marked by a vertical dash-dot line in each panel) occurs approximately 94 s later at about 18:51:34, as determined by Temmer et al. (2010). The time of the peak acceleration predicted by the EFle solution is remarkably consistent with the time determined by Temmer et al. (2010), who used the observation from the STEREO-B instruments to derive the kinematics. Their peak acceleration, however, appears to be slightly lower than our determined value by a visual inspection. Chen and Kunkel (2010), who used the images from STEREO-A and applied the same EF equations to model the data, obtained an acceleration peak slightly earlier than 18:50 UT (cf. Fig. 6 in their paper).

The SXR emission light curve - $I_{SXR}(t)$ - and the $d\Phi_p/dt$ profile predicted by the EFle solution for this event are compared in the left panel of Fig. 7. The two curves coincide with each other, and the timing of the peaks is in agreement. This consistency is visibly better than the results by Chen and Kunkel (2010) (cf. Fig. 6), who used the same model but only the data from STEREO-A instruments. The lack of the initial part of $d\Phi_p/dt$ reflects the lack of observation for the initial rising stage in our observations. We note that the decay phase of $I_{SXR}$ and that of $d\Phi_p/dt$ differ. Because the EF model does not include radiation or radiation-plasma interaction, it is beyond the scope of this paper to address the decay time scale. In addition, since the model does not include the particle energization/acceleration underlying the SXR emissions, it produces no prediction of the energy spectrum. It does, however, directly predict the temporal profile of the electromotive force [$\varepsilon(t) \equiv -(1/c)d\Phi_p(t)/dt$] that can accelerate particles. Thus, the comparisons here are between the predicted and observed temporal profiles excluding the details of the decay phase. The detailed manifestations and magnitude of $d\Phi_p/dt$ cannot be resolved or measured at this time, and must await future modeling and observations.

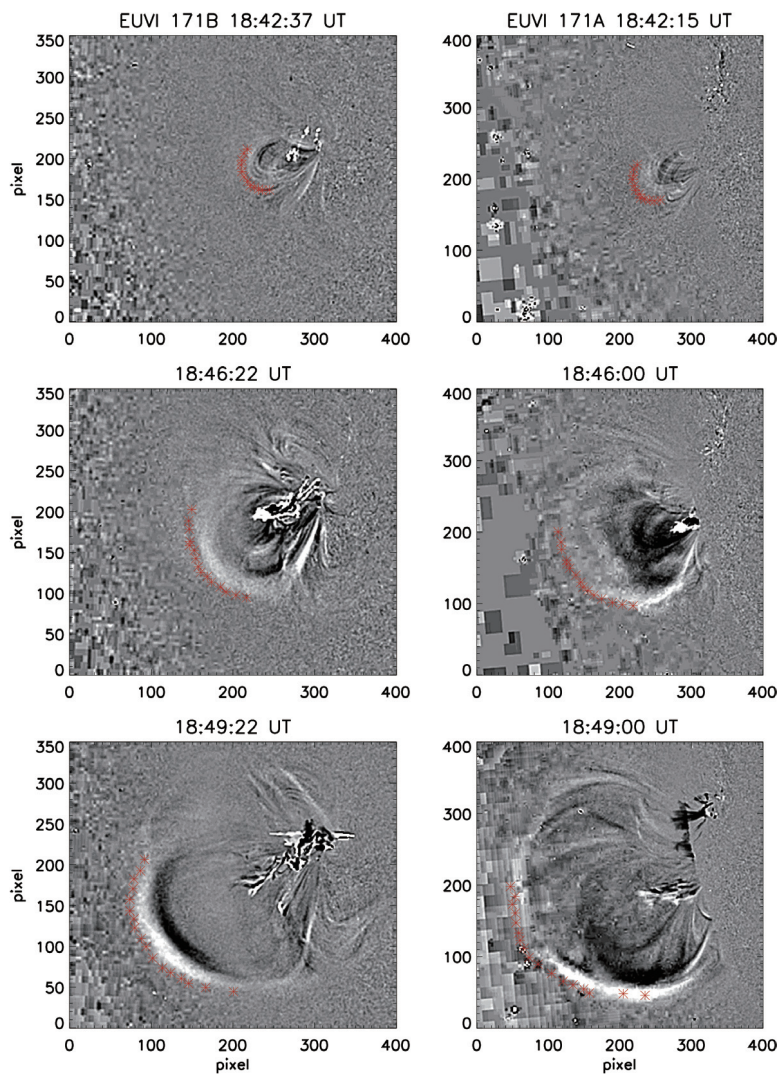

Fig. 3. The running-difference EUVI images of CME03. EUVI171-A and -B images are plotted in the right-hand-side (RHS) and left-hand-side (LHS) panels, respectively. The red stars mark the points along the coronal mass ejection (CME) front. The observation times are indicated above the corresponding panels. The white rising feature in the core of the CME is an erupting prominence.

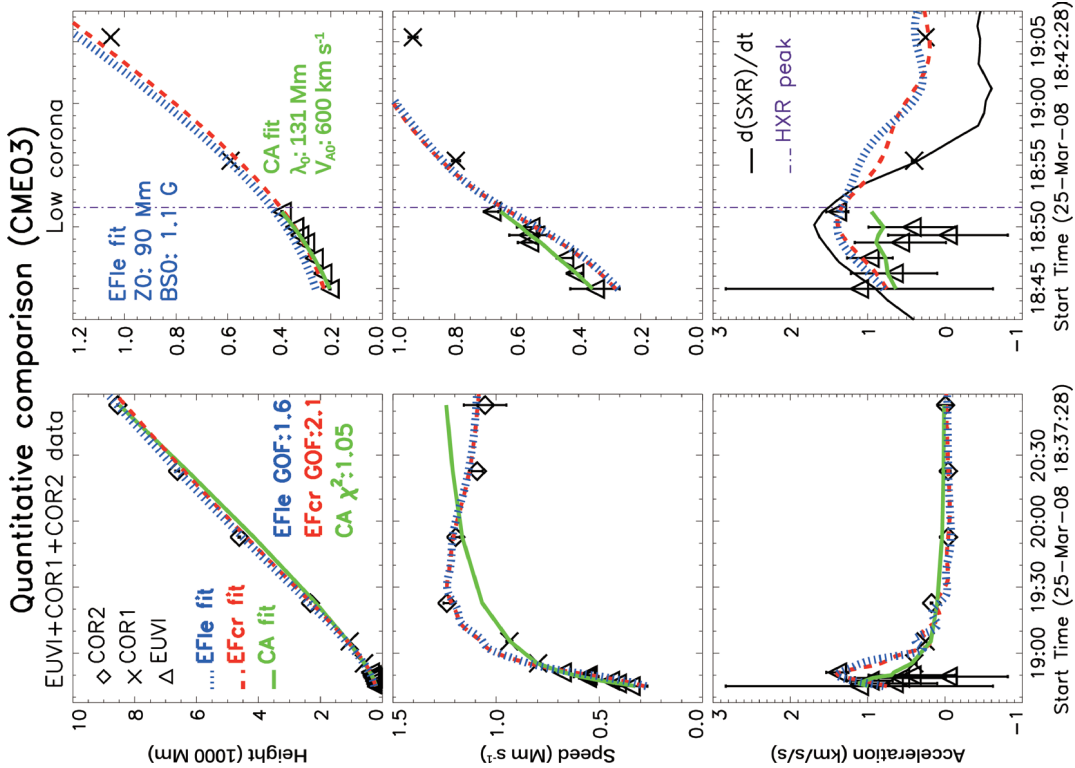

Fig. 5. The fitting results for CME03. The data and best-fit results are plotted in symbols and lines as indicated. The LHS column shows the complete profile while the RHS shows only the profile in the low corona to enhance visibility of the initial eruption. The dot-dash vertical line in the RHS column marks the time of HXR emission peak. The solid curve in the lower right panel is the time derivative of SXR light curve. EF fit: Z0 is the initial equilibrium height from the bottom of corona, and BS0 is the external coronal field perpendicular to the flux rope at Z0. CA fit: $\lambda_0$ is the critical height, and $V_{A0}$ is the Alfvén speed at $\lambda_0$.

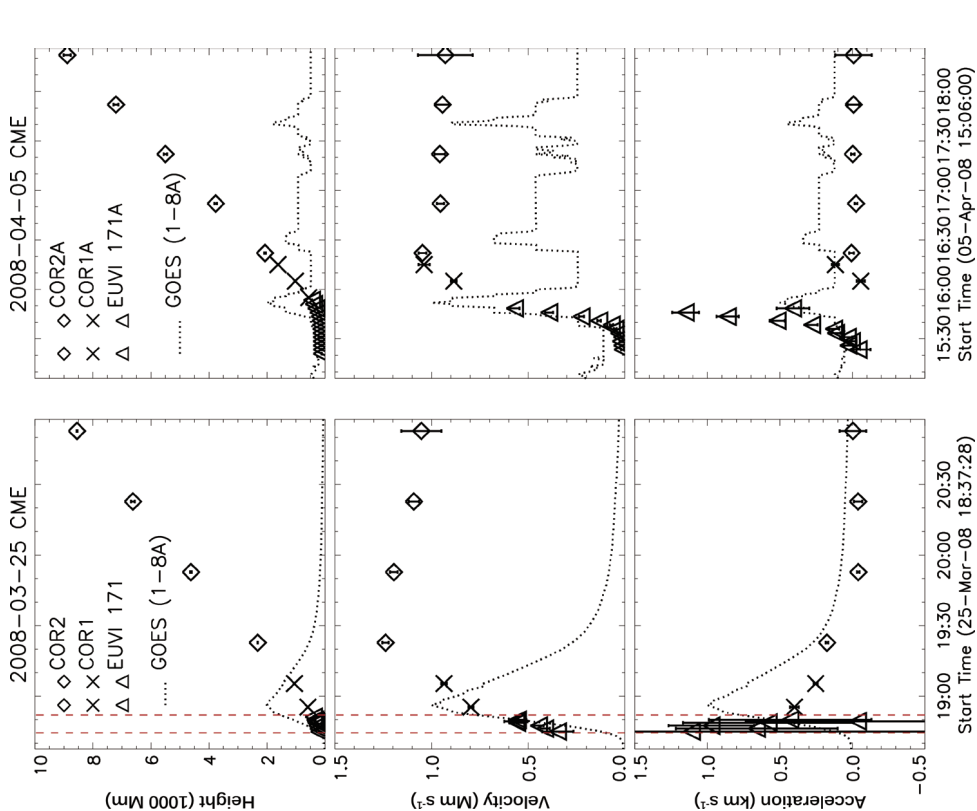

Fig. 4. (Time in UT) The observationally derived kinematics of the CMEs and associated Geostationary Operational Environmental Satellite soft X-ray (GOES SXR) light curves. The CME03 and CME04 results are presented in the LHS and RHS columns, respectively. The SXR magnitude has been scaled to enhance the visibility. The red dashed lines mark the beginning and the peak times of the hard X-ray (HXR) emissions associated with CME03.

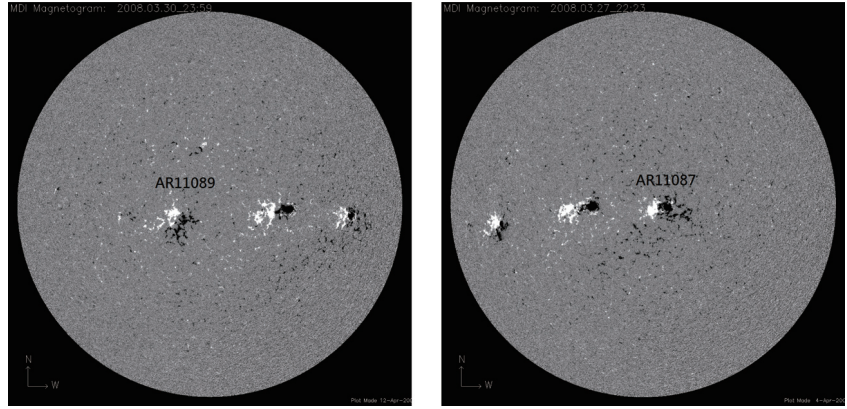

Fig. 6. The MDI magnetograms of AR10989 (left panel) and AR10987 (right panel) when they were located at the disk center. AR10989 is the source region of the CME erupted on 25 March 2008 (CME03), and AR10987 is the source region for the 5 April 2008 event (CME04).

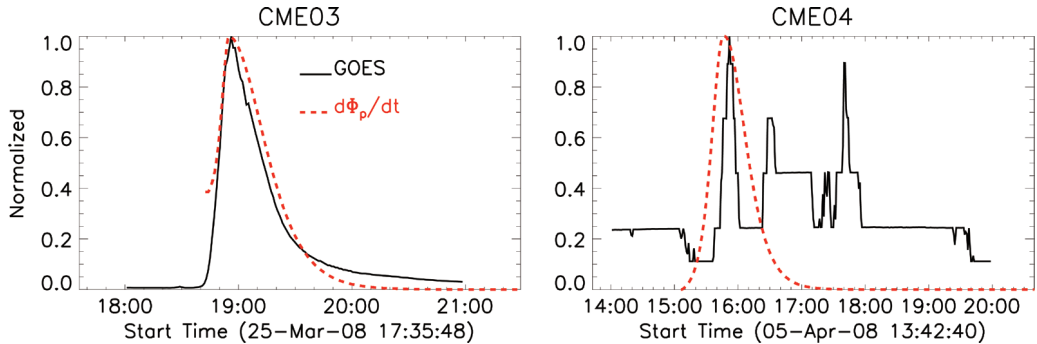

Fig. 7. (Time in UT) Comparison of the GOES SXR profile (solid black line) and the poloidal magnetic flux injection rate, $d\Phi_p/dt$, (red dashed line) predicted by the EF best-fit. The CME03 and CME04 results are shown in the left and right panels, respectively. All magnitudes have been normalized to allow easy comparison of the shapes.

**4.2 5 April 2008 Event (CME04)**

CME04 was seen as a west limb event from the viewpoint of STEREO-A but was occulted by the solar disk in the EUVI images as seen from STEREO-B. Therefore, the kinematics for this event was derived solely from the STEREO-A instruments. The motion of the CME was determined by tracking the CME front in each image. The procedure was repeated ten times for each image to obtain an average height. For a CME radially propagating from the eruption site, observed elongation is converted to distance from Sun center using the trigonometric formula (e.g., Chen and Kunkel 2010):

$$h = R_{obs} \frac{\sin\alpha}{\sin(\alpha + \mu)} \quad (5)$$

Where $h$ is the corrected height from the solar center, $R_{obs} = 214R_\odot$ is the observer-Sun distance, $\alpha$ is the elongation, and $\mu \equiv \cos^{-1}(\cos\Phi \cos\theta)$, where $\theta$ and $\Phi$ are the heliocentric longitude and latitude of the source region from the STEREO-A viewpoint. As described in section 4.1, the observational error considered was either the standard deviation of the ten trials or 1% of the determined height, whichever is larger. The observational velocity, acceleration, and their respective errors were subsequently propagated in the standard way as described in section 4.1.

The evolution of the source region during the CME04 process is shown in the LHS column of Fig. 8, and the running difference images of the respective images are plotted in the RHS column to show the CME evolution at the same time. The middle row reveals the brightening/appearance of a small-scale arcade at the side of one foot point of the CME when the CME reached the EUVI field of view boundary. The last row shows the brightening of several footpoints in the source region after the CME flew off. These phenomena indicate the occurrence of multiple heading events/processes associated with the CME.

The SXR light curve for this event is plotted in the right panel of Fig. 7, which shows multiple emission peaks. Because the source region was occulted by the west limb of the Sun from the GOES viewing angle, the SXR flare data represent emissions from the top part of the flaring region in the corona. As a result, the onset time inferred from this data is later than the actual SXR onset by an unknown amount and for each peak, the duration is shorter than the actual duration including the occulted emissions, for example, as measured by the full width at half maximum (FWHM). Each peak of the SXR light curve, however, should correspond to the actual peak in the SXR emissions of the flare event. These expectations may be invalid, however, if the intensity variation in the occulted low corona is significantly different from that above the limb from the GOES vantage.

Comparing the timing of these enhancements with the times indicated in Fig. 8, we found that the first peak (~15:51 UT) coincided with the brightening of the side arcade and the second one (~16:30 UT) with the brightening of the foot points, indicating a possible association between the two SXR emission enhancements and the brightening at the source region.

The kinematics of this CME and the associated SXR light curve (dotted) are shown in the right column of Fig. 4. In contrast to CME03, CME04 was indeed observed from its initial equilibrium state, as evidenced by the near-zero initial speed. The velocity of the CME began to increase after ~15:30 UT, peaked around 16:20 UT, and then decreased to a near constant magnitude $\approx 1000$ km s$^{-1}$ at the end of the observation. The observed acceleration peaked around 15:42 UT at a magnitude of approximately 1.2 km s$^{-2}$. While the initial stage of the eruption was visible by STEREO-A, it was occulted by the solar limb from the viewpoint of GOES. Therefore, we cannot determine the precise time at which SXR emission began to increase. Nevertheless, Fig. 8 shows a small flaring arcade at the side of one of the CME foot points, instead of directly below the erupted CME flux rope, as illustrated in Fig. 1b and as reported by Lin et al. (2010) in one of their studied CMEs (17 December 2006). In these events, enhancements in EUV and SXR emissions are typically interpreted in the literature as a result of magnetic reconnection. If this is true, then the multiple brightenings suggest the occurrence of reconnection at multiple stages during the eruption. The data, however, contain no information to prove or disprove this interpretation.

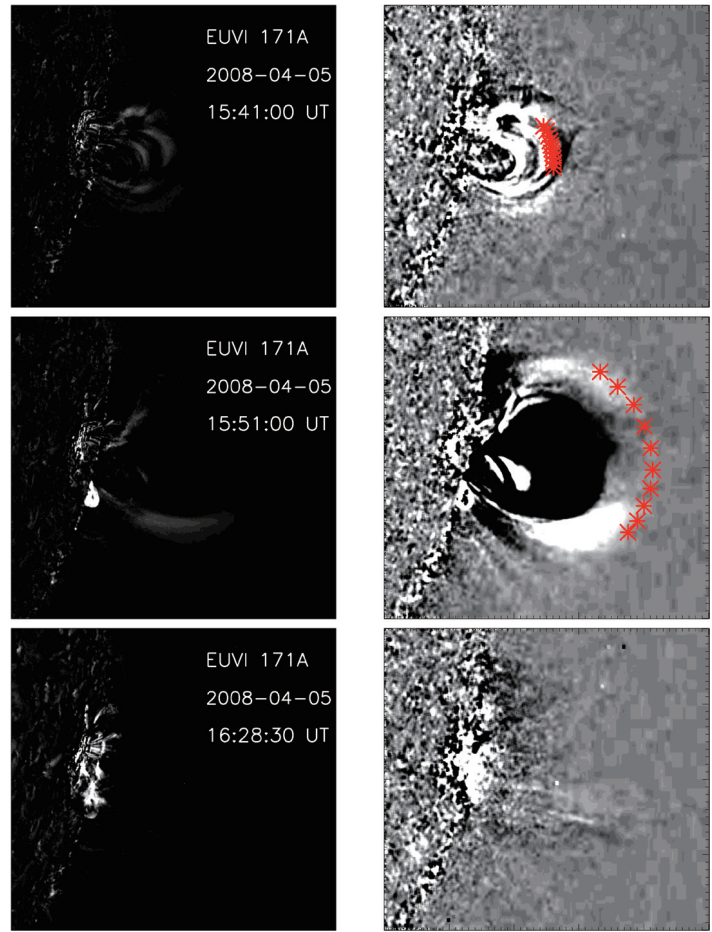

Fig. 8. Simultaneous evolution of CME04 and its source region at selected times, as indicated in the LHS. The LHS column shows the source region, and the RHS are the corresponding running difference images to show CME04. The red stars in the RHS mark the leading edge (LE) of the CME.

The quantitative examination of CME04 is presented in Fig. 9. The symbols and the line styles are the same as those in Fig. 5. The CA solution (green solid line) matches the eruption process up to the point of the observed peak velocity but did not undergo any deceleration: we found no CA solution that replicates the entire observed trajectory of the CME. The CA model predicted the critical height $\lambda_0$ for this event to be 58 Mm and the Alfvén speed at this height to be 500 km s$^{-1}$. These values translate to a foot point separation of $\approx$ 116 Mm, which is consistent with the size of AR10987 at photosphere ($\sim$150 Mm), and an ambient magnetic field strength of $\approx$ 18 G for a plasma density of $\approx 10^{-4}$ kg km$^{-3}$. For completeness, we applied the same procedure as described in section 4.1 to estimate the magnetic reconnection rate based on the CA solution for this event. By setting the magnetic field of the source region to be 80 G, we obtained an estimated $M_A \approx 0.025 - 0.03$ for this event. For both CME03 and CME04, the best-fit solutions of the CA model predict specific reconnection rates required by the observed height-time data. These rates cannot be validated by currently available data.

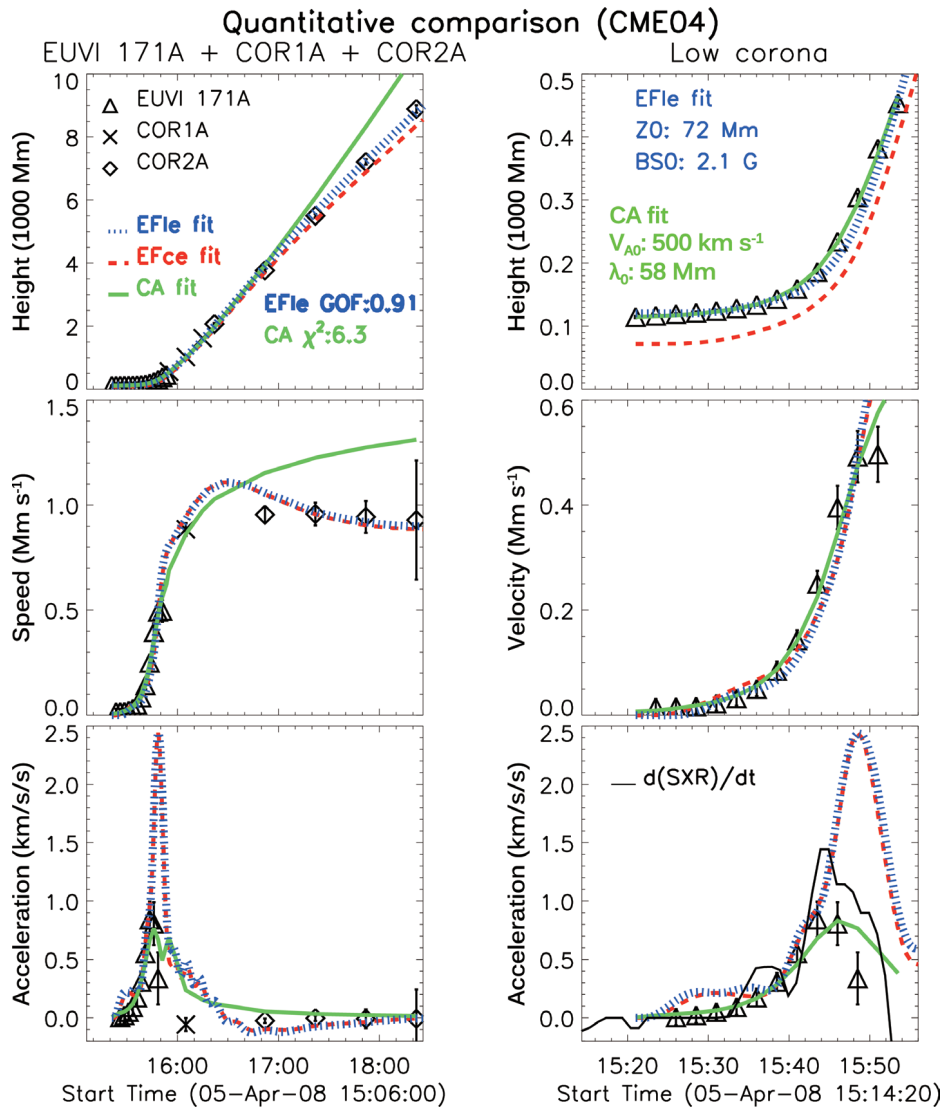

Fig. 9. EF and CA fitting results of CME04. The height, velocity and acceleration are plotted in the top, middle and bottom panels, respectively. The symbols and line styles are as described in Fig. 5.

The EFle solution (blue dotted line) closely matches the observed CME height-time data, except that the calculated main acceleration phase is sharply peaked with the maximum value approximately twice the observationally determined value. The observed acceleration profile may lack such a peak or may not resolve a narrow spike if one exists. This spike approximately coincides with the observed peak to within the time resolution but may be slightly later. The initial height and the foot-point separation obtained from the EFle solution are $Z_0 = 72$ Mm and $S_f = 140$ Mm, respectively. These valuables are consistent with the size of AR10987 at photosphere (~150 Mm), as estimated from the MDI magnetogram (cf. right panel in Fig. 6). The required $d\Phi_p/dt$ and the observed SXR of this event are compared in the right panel of Fig. 7. Since the source region of the event was behind the solar limb from the GOES view point, the instrument could only detect the emission after the SXR loops had grown sufficiently high to be seen over the limb, and, as a result, the earlier part of emission is missing in the GOES light curve. Thus, the onset of the SXR emissions detected above the limb should be later than the actual onset of SXR emissions including those behind the limb. This also reduces the width of the peak observed above the limb. The plot shows that the first SXR emission spike was detected after the peak of $d\Phi_p/dt$. The delay is approximately 10 min. The FWHM width of the observed spike is narrower than the width of $d\Phi_p/dt$. Another notable discrepancy between the predicted $d\Phi_p/dt$ and the SXR time curve is the lack of the second and third emission enhancements.

It is noted, however, that there is insufficient information to determine whether these two SXR emission peaks are indeed associated with CME04. It is not impossible that the post-CME brightening at the source region, which coincides with the second SXR peak, was due to processes unrelated to the acceleration of the specific CME structure. It is also noted that that both the CA and EF models are formulated in terms of one main energy release process, and neither model would predict multiple peaks in $I_{SXR}(t)$. It is not known, for example, how CME acceleration would be modified in the EF model if the flux injection function $d\Phi_p/dt$ had multiple peaks. It should be kept in mind that the EF model does not include any particle acceleration or radiation mechanisms so that the comparison of $d\Phi_p/dt$ and the SXR light curve is physically meaningful insofar as the temporal profiles of the EMF $\varepsilon(t) \equiv -(1/c)d\Phi_p(t)/dt$ and $I_{SXR}(t)$ are concerned. This also means that the theory does not describe the decay phase of the flare.

Antiochos et al. (1999) developed the Breakout model in which an erupting flux rope is formed in a multi-polar arcade system, consisting of a central arcade, two side arcades and an overlying arcade. The central arcade is what would become the erupting flux rope. The eruption process starts with the first reconnection between the central and the overlying arcade, followed by the second reconnections in the CS formed below the erupting flux rope. The last set of reconnections is between the side arcades to reform and restore the magnetic field in the low corona.

The small, flaring side arcade after the eruption and the brightening of multiple foot points at the source region may indicate that the above breakout scenario is at play. The multiple SXR emission enhancements can be qualitatively explained by the multiple sets of reconnection. However, since neither the simulation results nor predicted kinematic profiles of the model for this event are currently available, we cannot assess the physical consistency between the model and the observation to test the validity of the scenario. Nevertheless, there is no data to establish that the apparent side arcade is physically connected to the eruption of the observed CME.

## 5. SUMMARY

We compared two CME models, the EF and CA models, with the two most prominent CMEs observed during the WHI. The objective is to understand the CME driving mechanisms and test the two distinct theoretical processes represented by the models.

The EF model is formulated under the ideal MHD framework. The mechanism to drive an initial equilibrium flux rope into eruption is an increase in its poloidal magnetic flux. The scenario proposed by the CA model is that a magnetic arcade/flux rope undergoes quasi-static evolution due to magnetic footpoint motions in the photospheric boundary surface and catastrophically loses equilibrium via an ideal MHD process. Subsequently, magnetic reconnection is invoked to convert the overlying coronal magnetic flux into the flux-rope poloidal flux, accelerating the flux rope (a CME). As discussed in section 2.3, the expression for the rate of increase in the poloidal flux in the EF model is mathematically equivalent to the expression for the magnetic reconnection rate in the CA model. In the EF model, the mechanism of poloidal flux injection is not specified, while in the CA model, the reconnection rate is either prescribed or numerical (in simulations). These are two fundamental physical questions that are still open.

Because the CA and EF models use distinctly different physics, the comparison of these two models with the same CME height-time data allows one to better understand how the forces responsible for the observed CME acceleration operate. To the extent that both models use the Lorentz force ($J \times B$) as the primary driving force, one of the important differences found in the model acceleration profiles can be attributed to to the different geometries: the CA model uses a linear 2D geometry while the EF model uses a toroidal flux rope with stationary footpoints, a fully 3D geometry. In the EF model, the main driving force is the Lorentz hoop force, which is determined by the major radial curvature of the current-carrying toroidal plasma. In particular, the acceleration profile critically depends on the footpoint separation distance $S_f$, and the expanding CME flux rope experiences a significant drag force as the expansion speed increases, which can cause the flux rope to decelerate. In the CA model, this force is identically zero and with no stationary photospheric footpoints. In addition, there is no drag force. This is consistent with the CA solutions exhibiting no deceleration. One important general conclusion suggested by this work is that it is necessary to compare model results with the observed CME trajectories in their entirety: it is often possible to fit a small segment of an observed trajectory, but it is much more difficult to fit a trajectory in its entirety that can exhibit non-trivial dynamical effects.

The two selected events were launched on 25 March 2008 (CME03) from AR10989 and 5 April 2008 (CME04) from AR10987, respectively. CME03 was associated with an M 1.7 GOES class solar flare and a filament eruption. CME04 was launched behind the edge of solar limb. After the eruption, a bright side arcade appeared, and, approximately 50 minutes later, several foot points brightened at the source region. The timing of these brightenings coincided with multiple SXR emission enhancements detected in the GOES 1 - 8°A channel.

The kinematics of CME03 was computed using a stereoscopic reconstruction based on the data from both STEREO-A and -B instruments. The kinematics of CME04 was derived exclusively from the STEREO-A observation because most of the eruption process of the event was occulted by the solar limb from the STEREO-B perspective.

Our examination consists of a quantitative analysis, in which three theoretical CME trajectories were matched to those of corresponding CME features. Using the solutions that best match the kinematics of these features, theoretical values of certain observed properties are calculated and are compared with the corresponding observables. These quantities include the flux-rope footpoint separation distance $S_f$, the temporal profile of the SXR light curve $I_{SXR}(t)$, and the HXR emission profile.

For the 25 March 2008 (CME03) event, our analysis shows that the predictions from the EF model are consistent with the observed kinematic profile and the GOES SXR light curve. The foot point separation predicted by the model is in agreement with the source region size within 20%. The consistency indicates that the eruption of CME03 can be explained by the EF model. Although the CA model can be tuned to produce a kinematic profile that matches the observed rise phase of the initial acceleration, we found no single initial-value solution that can match the initial acceleration and subsequent deceleration.

For the 5 April 2008 (CME04) event, both EF and CA models can be tuned to produce kinematic profiles that are quantitatively consistent with the observation although the CA solution again fails to exhibit the observed deceleration following the initial acceleration. The footpoint separations predicted by both models are consistent with the source region size. The EF model also produced a $d\Phi_p/dt$ profile that coincides with the first SXR emission peak of this event. However, neither model can completely explain the observed post-eruption arcades, multiple brightenings and multiple SXR emission enhancements. While the data contain no information to show that the observed multiple SXR enhancements are or are not connected to the CME eruption, neither model treats multiple energy injection. If they are indeed integrally connected to the CME eruption, the CME04 event may qualitatively resemble the scenario envisioned by the breakout model (Antiochos et al. 1999), which calls for magnetic reconnection at different stages of an eruption. A quantitative comparison between a BO simulation of specific events with multiple brightenings and the determination of the physical reconnection rates in the coronal reconnection sites will be required to test this possibility.

**Acknowledgements** This work is funded by the NSC of ROC under grant NSC99-2112-M-008-019-MY3, NSC102-2112-M-008-018, and the MOE grant "Aim for the Top University" to the National Central University. JC is supported by the Naval Research Laboratory Base Program. The SECCHI data are produced by an international consortium of the NRL, LMSAL, and NASA GSFC (USA), RAL and Univ. Bham (UK), MPS (Germany), CSL (Belgium), IOTA and IAS (France). CHL wishes to thank Jun Lin, Angelos Vourlidas and Lou-Chuang Lee for helpful inputs and suggestion.